 \documentclass[preprint2]{aastex}

\shorttitle{Hot Halos in Field Galaxies}
\shortauthors{Mulchaey et al.}

\begin{document}


\title{Hot Gas Halos in Early-Type Field Galaxies}

\author{John S. Mulchaey} 
\affil{The Observatories of the Carnegie Institution of Science, 813 Santa Barbara St.,
    Pasadena, CA 91101, U.S.A.}
\email{mulchaey@obs.carnegiescience.edu}

\and

\author{Tesla E. Jeltema\altaffilmark{1}}
\affil{UCO/Lick Observatories, 1156 High St., Santa Cruz, CA 95064, U.S.A.}
\email{tesla@ucolick.org}
\altaffiltext{1}{Morrison Fellow}

\begin{abstract}
We use {\it Chandra} and {\it XMM-Newton} to study the hot gas content
in a sample of field early-type galaxies.  We find that the L$_{\rm
X}$--L$_{\rm K}$ relationship is steeper for field galaxies than for
comparable galaxies in groups and clusters.  The low hot gas content
of field galaxies with L$_{\rm K} \lessapprox L_{\star}$ suggests that
internal processes such as supernovae driven winds or AGN feedback
expel hot gas from low mass galaxies. Such mechanisms may be less
effective in groups and clusters where the presence of an intragroup
or intracluster medium can confine outflowing material.  In addition,
galaxies in groups and clusters may be able to accrete gas from the
ambient medium.  While there is a population of L$_{\rm K} \lessapprox
L_{\star}$ galaxies in groups and clusters that retain hot gas halos,
some galaxies in these rich environments, including brighter galaxies,
are largely devoid of hot gas. In these cases, the hot gas halos have
likely been removed via ram pressure stripping.  This suggests a very
complex interplay between the intragroup/intracluster medium and hot
gas halos of galaxies in rich environments with the ambient medium
helping to confine or even enhance the halos in some cases and acting
to remove gas in others.  In contrast, the hot gas content of more
isolated galaxies is largely a function of the mass of the galaxy,
with more massive galaxies able to maintain their halos, while in
lower mass systems the hot gas escapes in outflowing winds.
\end{abstract}


\section{Introduction}

It has been known since the mid-1980's that hot gas halos are common
in early-type galaxies \citep{forman85}.  From the earliest studies
with {\it Einstein}, it was apparent that the X-ray luminosity of
early-type galaxies is correlated with the stellar luminosity
\citep{forman85,tf85,c87}.  The correlation between these two
quantities suggests that the origin of the hot gas must be related to
the stellar content of the galaxy. It is generally believed that the
hot gas originates from stellar mass lost from evolved stars and
planetary nebula \citep{m90,mb03,BP09}. However, the scatter in the
L$_{\rm X}$-L$_{\rm B}$ relationship is very large. Since the B band
can be strongly affected by both recent star formation and dust, it
may not be a good measure of the true stellar content of a galaxy in
some cases.  The large scatter in the relationship remains, however,
when L$_{\rm K}$ is used as a proxy for stellar light \citep{ellis06},
suggesting the scatter is dominated by variations in the X-ray
properties of galaxies. The large range in X-ray luminosity for a
given stellar luminosity could be due to either intrinsic differences
in galaxy properties or environmental effects.

There has been considerable effort by the astronomical community to
understand the scatter in the L$_{\rm X}$-L$_{\rm B}$ and L$_{\rm
X}$-L$_{\rm K}$ relationships with different authors often reaching
opposing conclusions. For example, some authors have found evidence
for the X-ray luminosities of early-type galaxies to vary with
environment \citep{ws91,brown00}, while others find no such trend
\citep{o01,h01,ellis06}.  One problem that plagued these earlier studies
was the inability to cleanly separate out the thermal emission in
galaxies from other contributions to the X-ray emission. In
particular, the contribution of X-ray binaries and an active galactic
nucleus (AGN) can be substantial in some galaxies. The broader
bandpasses and superior spatial resolution of {\it Chandra} and {\it
XMM-Newton} allow a much cleaner measurement of the thermal gas
component than was possible with earlier telescopes \citep{kf03,diehl07}.

The superb spatial resolution of {\it Chandra} is also important
because it allows one to separate out an individual galaxy's hot gas
halo from the more extended intragroup or cluster medium. Recent
studies of early-type galaxies in groups and clusters indicate that a
large fraction of such galaxies retain their hot gas halos even in
these dense environments \citep{sun07,jbm08,sun09}.  The presence of
hot gas halos in group and cluster galaxies has important consequences
for galaxy evolution.  In nearly all models of galaxy formation, the
condensation of hot halo gas is a primary driver for the build up of
massive galaxies \citep{wf91,cole00,bower06}.  Since the pioneering
work of \citet{wf91}, it has generally been assumed that a galaxy's
hot gas halo is stripped completely when a galaxy enters a group or
cluster. With the hot gas halo removed, there is no new source for gas
and the star formation rate quickly declines and the galaxy becomes
red.  The {\it Chandra} observations of hot gas halos in groups and
clusters demonstrate that the assumption of complete stripping is
overly simple. Some authors have started to incorporate more
sophisticated stripping prescriptions in to the semi-analytic models
and a more realistic treatment of the hot gas appears to alleviate
some problems that were present in the earlier versions of the models
\citep{kang08,mccarthy08,font08}.

To quantify the importance of stripping in groups and clusters, the
properties of ellipticals in these rich environments must be compared
to the properties of galaxies in environments where stripping is not
important, i.e. the field.  \citet{jbm08} attempted to make this
comparison and found that group and cluster galaxies appear to be
X-ray-faint compared to field galaxies. However, this result was based
on {\it ROSAT} observations of field galaxies, so the contribution of
X-ray binaries and AGN had to be estimated.  In this {\it Letter}, we
use {\it Chandra} and {\it XMM-Newton} observations of field
early-type galaxies to cleanly measure their hot gas content for the
first time.  We then compare the X-ray properties of these field
galaxies to similar galaxies in groups and clusters to study how
environment impacts hot halos in galaxies.  We adopt H$_{0}$=70 km
s$^{-1}$ Mpc$^{-1}$, $\Omega$$_{\rm M}$ = 0.3 and
$\Omega$$_{\Lambda}$= 0.7 throughout this {\it Letter}.

\section{The Sample}

Our goal is to study the hot gas content in a sample of early-type
galaxies outside of rich groups and clusters.  We derive our sample
from published catalogs of nearby field early-type galaxies
\citep{colbert01,reda04,ellis06,m09}.  To verify isolation, we used
NED to eliminate galaxies that are in cataloged groups and clusters
or in close pairs.  We also examined the field around each galaxy
using the POSS to verify that there were no luminous neighbors
uncataloged in NED. Adopting these criteria guarantee that these
galaxies lie in low density environments quite unlike the X-ray
luminous groups and clusters used in earlier studies
\citep{sun07,jbm08}.
%Finally, we
%eliminated well-known fossil groups from our sample as
%these galaxies are similar to central brightest group and cluster
%galaxies which were not considered in the previous studies \citep{sun07,jbm08}.
To ensure that the targets would be bright enough for study with {\it
Chandra} and {\it XMM-Newton}, we restricted our sample to galaxies
with redshifts less than z=0.03.  Finally, we restrict our analysis to
galaxies more luminous than Log L$_{\rm K}$ = 10.5 to allow a direct
comparison to the previously studied group and cluster samples.
Applying these criteria results in a sample of 74 nearby field
early-type galaxies.  We observed five of these galaxies with {\it
Chandra} and four with {\it XMM-Newton}. An additional eighteen
galaxies have observations available in the {\it Chandra} and {\it
XMM-Newton} archives.  Several of the {\it XMM-Newton} observations
suffered from severe flaring resulting in very short effective
exposure times.  Removing these objects results in a final sample of
23 galaxies with sufficient X-ray observations for our purposes (Table 1)

The {\it Chandra} data were prepared using standard reduction
processing in CIAO 4.1.2 and CALDB 4.1.4 following the method
described in \citet{jbm08}, while the {\it XMM-Newton} data were
reduced using SAS 8.0.0 following the methods outlined in
\citet{jeltema06}.  For both the {\it Chandra} and {\it XMM-Newton}
data, we extract radial profiles to determine the extent of the
X-ray emission.  All of the galaxies in our sample are detected as
extended sources.  Spectra were then extracted using a circular
aperture with radius equal to the maximum extent of the X-ray
emission.  Separate spectra were extracted for each of the three {\it
XMM-Newton} EPIC cameras. A local background was determined in each
case from a nearby source-free region.

All of the galaxies in our sample have enough counts to allow us to
extract an X-ray spectrum.  The spectral analysis was performed using
XSPEC 12.5.1. We use an identical fitting procedure to that used in
\citet{jbm08} to allow a direct comparison between the properties of
the hot gas in our field sample and their group sample. The method
adopted by \cite{jbm08} is virtually identical to the technique used
by \cite{sun07} for rich clusters.  The spectra were fit in the 0.5--7
keV band to allow a better determination of the contribution of X-ray
binaries or a potential AGN component.  For the spectral fits we adopt
a two component spectral model consisting of a thermal MEKAL model and
a powerlaw. The column density was fixed at the Galactic value and the
gas metallicity fixed at a value of 0.8 solar. For sources with
sufficient counts, we allow both the temperature and the photon index
to vary. In cases where both components could not be constrained, we
fix the photon index at 1.7. To compare with previous work, we report
the luminosity of any detected thermal component in the 0.5-2 keV
band.  Errors on the luminosity were determined using Monte Carlo
Markov chains.
%We adopt
%the median luminosity of the chain as the final thermal
%luminosity. 
For a few galaxies, a single powerlaw with index 1.7 provides an
adequate fit to the X-ray spectrum.  In these cases, we derive an
upper limit on any thermal component by first fitting the source
spectrum to a power-law model and then adding a thermal component with
the temperature fixed at 0.7 keV.  The upper limit on the luminosity
is set to the 3 $\sigma$ upper limit of this thermal component.

%Several of the galaxies in our sample contain several thousand counts in
%the {\it Chandra} and {\it XMM-Newton} observations. We will present a more 
%detailed study of the hot gas properties of these galaxies in a future 
%paper.

\section{The L$_{\rm X}-L_{\rm K}$ Relationship for Field Galaxies}

As we are able to extract a spectrum for every galaxy in our sample,
we are able to measure the luminosity of the thermal component or
calculate an upper limit for the thermal gas in each case.  Figure 1
shows the L$_{\rm X}-L_{\rm K}$ relation for our sample, where the
L$_{\rm X}$ measurement is for the thermal gas only. For comparison,
we plot the same quantities for early-type galaxies from the cluster
sample of \citet{sun07} and the group sample of \citet{jbm08}. We do
not plot the upper limits for the group and cluster sample because in
many cases the measurements represent the upper limit on the total
X-ray luminosity and not the thermal component only.  Using the
bisector modification to the BCES method in \citet{ab96}, we derive
the best-fit to the field galaxy L$_{\rm X}-L_{\rm K}$ relation:

\centerline{log(L$_{\rm 0.5-2 keV}$) = 38.90$\pm{0.18}$ + 3.92$\pm{0.39}$ log(L$_{\rm K}$/10$^{11}$L$_{\rm K,\odot}$).}

\noindent{ The upper limits have not been used in the fit. Including
the upper limits results in a slightly steeper relationship.  The best
fit to the field galaxies is given by the dashed line in Figure 1,
while the best fit to the combined group and cluster sample is shown
by the solid line.}

As can be seen from the Figure, a thermal component is detected in all
field galaxies with K-band luminosities greater than Log L$_{\rm K}$
$\approx$ 11.2.  The X-ray luminosities of these galaxies are
comparable to or slightly more luminous than their counterparts in
groups and clusters. However, for galaxies with K-band luminosities
less than Log L$_{\rm K}$ $\approx$ 11.2, the field galaxies appear to
deviate from the relationship found for the detected galaxies in
groups and clusters. In particular, most of the field galaxies have
lower thermal luminosities than galaxies detected in groups and
clusters.  The general behavior described above leads to a much
steeper L$_{\rm X}-L_{\rm K}$ relationship for field galaxies than for
comparable galaxies in richer environments.  For comparison, the slope
of the L$_{\rm X}-L_{\rm K}$ relationship for group and cluster
galaxies is $1.86 \pm 0.23$ \citep{jbm08}, which is different than the
slope we derive for the field galaxies at greater than 5 $\sigma$
significance.  Therefore, our study appears to confirm that
environment contributes significantly to the scatter seen in the
L$_{\rm X}-L_{\rm B}$ and L$_{\rm X}-L_{\rm K}$ relationships. It is
interesting to note that the stellar luminosity where the field
galaxies transition from having detected hot halos to having little or
no hot gas (Log L$_{\rm K}$ $\approx$ 11.2) is close to the K-band
value of L$_{\star}$ (L$_{K \star}$ = 10$^{11.08}$ L$_{\rm K,\odot}$
\cite{k01}). Therefore, even fairly massive field galaxies do not have
significant hot gas halos.

The field galaxy L$_{\rm X}-L_{\rm K}$ relation presented here based
on {\it Chandra} and {\it XMM-Newton} data differs significantly from
previous relationships based on {\it ROSAT} data, where all of the
detected X-ray emission was assumed to be thermal
\citep{ellis06,jbm08}, having both a lower normalization and a steeper
slope.  This difference highlights the importance of separating out
the different contributions to the X-ray emission.  While there is
little doubt that including a power law component in the spectral fits
helps account for the contribution of X-ray binaries, there may still
be some contribution to the X-ray emission from other stellar sources.
In particular, the contribution of cataclysmic variables (CVs) and
coronally active binaries (ABs) could be significant in the lower
luminosity sources and it is difficult to spectrally separate this
emission since its spectrum can be approximated by the combination of
a thermal plasma and a powerlaw \citep{trinch08,rev08}.  To estimate
the importance of this effect, we have plotted the expected X-ray
luminosity for these stellar populations as a function of the K-band
luminosity in Figure 1 using the relationship derived by \citet{rev08}
for early-type galaxies. As can be seen from Figure 1, the
contribution of these stars can potentially account for all of the
X-ray emission attributed to thermal gas in the lowest luminosity
field galaxy we detect, NGC 3115.  For the rest of the detected field
galaxies, the contribution of CVs and ABs is unlikely to be
significant.  Given that the NGC 3115 thermal measurement may not be
secure, we fit the L$_{\rm X}-L_{\rm K}$ relationship with this galaxy
removed.  The resulting slope (3.54$\pm{0.36}$) is still significantly
steeper than the group and cluster relationship.

\section{Discussion}

In the previous section we show that the L$_{\rm X}-L_{\rm K}$
relation is steeper for field galaxies than for galaxies in groups and
clusters. This result has important implications for the role of
environment on the hot gas halos of galaxies. In particular, our study
suggests there is a population of early-type galaxies with L$_{\rm K}$
$\lessapprox$ L$_{\star}$ in groups and clusters that retain
substantial hot gas reservoirs, while their counterparts in the field
are mostly devoid of gas.

The lack of a substantial hot gas component in field galaxies with
L$_{\rm K}$ $\lessapprox$ L$_{\star}$ could reflect a fundamental
difference in the global properties of field and group/cluster
early-type galaxies.  For example, previous studies of early-type
galaxies have suggested a possible trend between the X-ray luminosity
and the age of a galaxy estimated from dynamical or spectroscopic
indicators \citep{o01}, with luminous X-ray emission apparently
restricted to galaxies with ages greater than a few Gyrs.  These
observations are consistent with a scenario where hot gas is initially
removed during major mergers and the hot gas halos take several
gigayears to build up \citep{cox06}.  The difference between the field
galaxies and those in richer environments could therefore reflect a
difference in age.  In fact, there is some indication from simulations
that isolated ellipticals should be on average younger than their
counterparts in groups and clusters \citep{sami10}.  However,
\citet{reda04} have studied the stellar populations in several of the
field galaxies in our sample and found that the bulk of the stars in
these galaxies are very old. In fact, for the galaxies they studied,
\citet{reda04} found that the formation epoch of field and cluster
ellipticals appears similar.  This suggests that age differences are
unlikely to be the explanation for the observed differences in the hot
gas content.

Another possibility is that field galaxies lack substantial dark
matter halos and are therefore unable to keep X-ray halos
\citep{o04}. Minimal dark matter halos have been implied for some
elliptical galaxies from kinematic studies of planetary nebula at
large radii \citep{r03,d07}.  However, the low velocity dispersions
derived from the planetary nebula could be due to halo stars on radial
orbits and not small dark matter halos \citep{dekel05}.  To explain
the observed X-ray/optical relationships, dark matter halos of low
L$_{\rm K}$ ellipticals in groups and clusters would need to be more
substantial than their counterparts in the field.  However, one might
naively expect the dark matter halos of ellipticals in groups and
clusters to be reduced relative to those in the field since tidal
stripping is much more likely in these denser environments.  Thus, the
differences in the scaling relations of field and group/cluster
galaxies most likely do not reflect differences in the dark matter
halos.

In addition to potential intrinsic differences between field and
group/cluster early-types, environmental processes could be important.
However, environmental processes that remove gas from galaxies (such
as ram pressure stripping) are likely only important in richer
environments, where there is a substantial intragroup or intracluster
medium \citep{km08,bekki09}. This suggests that internal processes
must be responsible for removing hot gas from low mass field
galaxies. Most likely gas has been expelled from these galaxies by
stellar winds or AGN feedback. Detailed studies of low X-ray
luminosity ellipticals have concluded that winds sustained by Type 1a
supernovae are likely the dominant mechanism by which galaxies lose
their hot gas, although AGN outbursts may also be important in some
cases \citep{david06,pel07,trinch08}.  In our field sample, there is
little evidence for significant AGN activity. Approximately half of
our sample galaxies are detected in radio continuum in the
NVSS \citep{condon98}, but in nearly all cases the emission is very
weak.  However, the weak radio emission in our field sample does not
necessarily mean that AGN feedback is not important in these systems,
since there is little correlation between 1.4 GHz radio luminosity and
disturbed X-ray morphologies (i.e. cavities) in many nearby
ellipticals \citep{dong10}.

While supernovae or AGN driven outflows can explain the low hot gas
content of field galaxies, such mechanisms may be less effective in
groups and clusters where the presence of an ambient medium may stifle
such winds \citep{babul92,babul99,brown98,brown00}.  In addition,
early-type galaxies in groups and clusters may be able to accrete gas
from the intragroup or intracluster medium \citep{bm98,bm99,brown00}.
The combination of these two effects likely accounts for the
population of low L$_{\rm K}$ group and cluster galaxies that still
contain significant amounts of hot gas.  There are several
observations that could help test the relative importance of gas
accretion from the ambient medium versus the suffocation of outflowing
winds. If accretion of gas is the dominant mechanism by which low
L$_{\rm K}$ cluster galaxies maintain halos, we might expect the
metallicity of the gas to be lower than if the gas is produced
internally in the galaxies.  Deeper X-ray observations of low L$_{\rm
K}$ group and cluster galaxies with halos should allow this test to be
performed.

We note that the L$_{\rm X}-L_{\rm K}$ relationships shown in Figure 1
were derived for galaxies with a detected hot gas halo.  Given that it
is more difficult to detect individual hot halos in groups and
clusters (because of the higher \lq\lq background\rq\rq \ from the
intragroup/intracluster medium), we likely
could not detect the very low L$_{\rm X}$ halos in these richer
environments that we detect in the field.  In fact, there are many
galaxies in groups and clusters where a thermal component has not been
detected \citep[see][]{sun07,jbm08} and in some cases the limits on
the hot gas luminosity would place these galaxies well-below the
group/cluster L$_{\rm X}-L_{\rm K}$ relationship shown in Figure 1.  This suggests
that although there is a population of low L$_{\rm K}$ galaxies in
groups and clusters that retain hot gas halos, there are other
galaxies, including brighter galaxies, in these rich environments that
have likely lost their hot gas halos to ram pressure stripping.  Our
study therefore suggests a very complex interplay between the
intragroup/intracluster medium and the hot gas halos of galaxies in
rich environments: the presence of an ambient medium can act to
maintain or even enhance a hot halo in some galaxies and remove halo
gas in other cases. In contrast, the hot gas content of more isolated
galaxies is largely a function of the mass of the galaxy, with more
massive galaxies able to maintain their halos, while the hot gas is
expelled in lower mass systems. To better understand the importance of
the various environmental processes at play in groups and clusters,
studies of how the properties of hot gas halos vary spatially in
groups and clusters would be valuable.

\acknowledgments

We acknowledge useful discussions with Arif Babul, Richard Bower, TJ
Cox, Ewan O'Sullivan, Jesper Rasmussen, David Wilman and Ann
Zabludoff.  JSM acknowledges support from Chandra grant
G09-0099A.

\begin{deluxetable}{cccccc}
\tablecaption{Early-Type Galaxy Field Sample}
\tablenum{1}
\tablehead{\colhead{Galaxy} & \colhead{z} & \colhead{log (L$_{\rm K}/L_{\rm K,\odot})$}& \colhead{Sample} & \colhead{Telescope}
& \colhead{log L$_{\rm X, 0.5-2 keV}$}} 
%% All data must appear between the \startdata and \enddata commands
\startdata
NGC57   & 0.018146 & 11.69  & E006 &  Chandra/XMM& 41.19$^{+0.02}_{-0.02}$ \\
NGC766  & 0.027032 & 11.61  & CMZ01 & Chandra& 41.03$^{+0.03}_{-0.04}$ \\
NGC821  & 0.005787 & 11.00  & R04/EO06 & Chandra& $<$38.31\\
NGC2325  & 0.007062& 11.18  & EO06 & XMM & 40.14$^{+0.04}_{-0.04}$\\
NGC2865 & 0.008763 & 11.14  & R04/EO06 & Chandra & 39.36$^{+0.07}_{-0.09}$ \\
NGC2954 & 0.012745 & 11.07  & M09 &  XMM & $<$39.14 \\
NGC2986 & 0.007679 & 11.35  & EO06 & Chandra & 40.29$^{+0.07}_{-0.09}$ \\
NGC3115 & 0.002212 & 10.96  & EO06 & Chandra &  38.32$^{+0.09}_{-0.11}$\\
NGC3209 & 0.020751 & 11.54  & CMZ01 & Chandra & 40.67$^{+0.05}_{-0.05}$\\
NGC3962 & 0.006054 & 11.13  & EO06 &  XMM& $<$39.47\\
IC2980  & 0.006978 & 10.72  & CMZ01 & XMM& $<$39.00\\
NGC4555 & 0.022292 & 11.67  & R04 & Chandra&  41.27$^{+0.04}_{-0.04}$\\
NGC4915 & 0.010117 & 11.13  & EO06 & Chandra &  39.35$^{+0.07}_{-0.08}$\\
NGC6127 & 0.016114 & 11.40  & EO06 & Chandra&  40.91$^{+0.03}_{-0.03}$\\
NGC6703 & 0.008209 & 11.17  & EO06 & Chandra& 39.67$^{+0.07}_{-0.08}$\\
IC4889  & 0.008586 & 11.26  & EO06 & XMM&  39.57$^{+0.09}_{-0.11}$\\
NGC7010 & 0.028306 & 11.64  & CMZ01 & XMM& 41.58$^{+0.02}_{-0.02}$ \\
NGC7029 & 0.009323 & 11.15  & EO06 & XMM& $<$38.86 \\
NGC7052 & 0.015584 & 11.60  & M09 & Chandra& 41.17$^{+0.02}_{-0.03}$\\
NGC7196 & 0.009737 & 11.30  & EO06 & Chandra/XMM& 40.59$^{+0.02}_{-0.02}$\\
NGC7507 & 0.005224 & 11.15  & EO06& XMM& $<$39.44\\
NGC7785 & 0.012702 & 11.46  & M09 & XMM& 40.93$^{+0.01}_{-0.01}$\\
NGC7796 & 0.010974 & 11.40  & R04/EO06& Chandra& 40.62$^{+0.01}_{-0.01}$\\
\enddata

\tablecomments{All redshifts are taken from NED. L$_{\rm K}$ values have been 
calculated from
total K$_{\rm short}$ magnitudes derived from the Two Micron All Sky Survey (2MASS; \cite{2mass}). The L$_{\rm X}$ values quoted are for the thermal 
component only.}

\tablerefs{CMZ01=Colbert, Mulchaey \& Zabludoff (2001); R04=Reda et al. (2004); EO06=Ellis \& O'Sullivan (2006); M09=Memola et al. (2009).}

%% Include any \tablenotetext{key}{text}, \tablerefs{ref list},
%% or \tablecomments{text} between the \enddata and
%% \end{deluxetable} commands

%% No \tablecomments indicated

%% No \tablerefs indicated

\end{deluxetable}

\begin{figure}
\epsscale{1.0}
\plotone{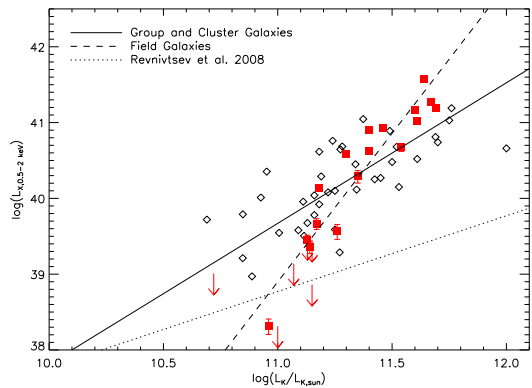}
\caption{L$_{\rm K}-L_{\rm X, 0.5-2 keV}$ relation for our early-type
field sample (red filled squares and upper limits) and early-type
galaxies in groups and clusters (open diamonds; taken from Sun et
al. 2007 and Jeltema et al. 2008).  All X-ray luminosities (including
upper limits) are for the thermal component only. The dashed line
represents the best fit to the detected field galaxies, while the
solid line gives the best fit relationship for the combined group and
cluster sample (see Jeltema et al. 2008). The dotted line shows the
estimated contribution to the X-ray emission from CVs and ABs based on
the relationship given in \citet{rev08}.  }
\end{figure}

%% The following command ends your manuscript. LaTeX will ignore any text

\end{document}